# Querying Large Physics Data Sets Over an Information Grid

Nigel Baker[1], Peter Brooks[1], Zsolt Kovacs[2], Jean-Marie Le Goff[2] & Richard McClatchey[1]

[1]Centre for Complex Cooperative Systems, UWE, Frenchay, Bristol BS16 1QY UK
Richard.McClatchey@cern.ch
[2]EP Division, CERN, Geneva, Switzerland
Jean-Marie.Le.Goff@cern.ch

**Abstract:** Optimising use of the Web (WWW) for LHC data analysis is a complex problem and illustrates the challenges arising from the integration of and computation across massive amounts of information distributed worldwide. Finding the right piece of information can, at times, be extremely time-consuming, if not impossible. So-called Grids have been proposed to facilitate LHC computing and many groups have embarked on studies of data replication, data migration and networking philosophies. Other aspects such as the role of 'middleware' for Grids are emerging as requiring research. This paper positions the need for appropriate middleware that enables users to resolve physics queries across massive data sets. It identifies the role of meta-data for query resolution and the importance of Information Grids for high-energy physics analysis rather than just Computational or Data Grids. This paper identifies software that is being implemented at CERN to enable the querying of very large collaborating HEP data-sets, initially being employed for the construction of CMS detectors.

**Introduction**

The fundamental problem the GRID research and development community is seeking to solve is how to co-ordinate distributed resources amongst a dynamic set of individuals and organizations in order to solve a common collaborative goal. The coordinated collection of resources, individuals and institutes is often called a virtual organization. The degree of distribution of an application that can run within such an organization can vary on a scale that runs from a centralized application that uses network resources but where control and data resides at one location to an application made up of a number of autonomous components that collaborate to meet some overall application goal. (one example of this is in the analysis of data for the resolution of HEP queries). In this later case the data and control of the application is fully distributed within the system, intelligent agents are exemplars of this type of collaborative behavior. The degrees of collaboration can be explained in terms of interoperability of the distributed components. Several levels of interoperability can be identified:-

- Network interfaces and protocols
- Component interfaces and data exchange
- Common services
- Meta model and knowledge

Level one interoperability is concerned with providing applications with a reliable error free data transport service across a diversity of physical networks. A TCP/IP protocol suite provides this type of interoperability. Level two is concerned with providing applications with a common programming model with well defined data types and remote call interactions. RPC toolkits and OMG CORBA (IDL, Orb and IIOP) are examples. But distributed components must conform to common service models such as security, location, naming, transaction, replication, group and notification services if they are to achieve any degree of collaboration.

Common services in the OMG OMA and Microsoft DCOM are examples at this level. The highest level of interoperability attempts to aspire to human cognitive levels of collaboration (the knowledge level). The argument is that it is only through an understanding of the meaning of data and the context in which it is used that higher levels of collaboration are possible. This requires common models and ontologies describing various contexts and domains such as manufacturing, business, HEP and telecommunications. This goal has spurned activities within all branches of computer science from distributed systems meta-modeling (OMG), federated databases, meta-computing, knowledge management and multi-agent systems.

**The GRID Collective Layer**

The GRID architecture in Figure 1 as defined by Foster et al [1] is very much an extensible framework within which to discuss interoperability issues faced by potential collaborative participants in a virtual organization (e.g. an LHC collaboration querying distributed physics data sets).

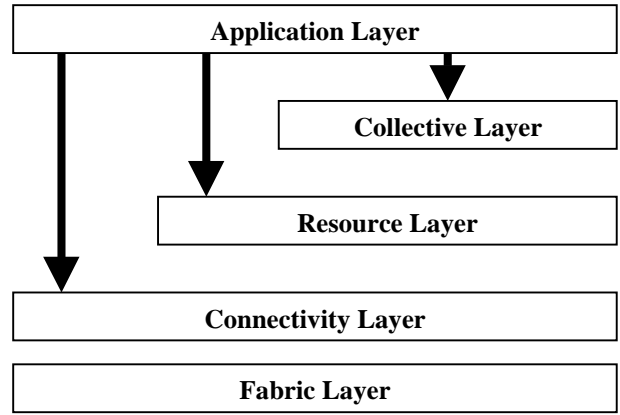

**Figure1**: Layered GRID Architecture

The collective layer is focused on interactions between collections of resources and is concerned with such things as directory services, replication services, discovery services, policy enforcement services, workflow management and collaborative frameworks. These collective services can be generic or specific to a particular application or domain. The Grid Computing Environment (GCE) research group of the Global Grid Forum (GGF [2]) are working at this level and aim to identify how developers are using the GRID and to contribute to the coherence and interoperability of frameworks, portals and problem solving environments (PSE's). A GCE extends the users desktop by providing tools that access the Grid services. Figure 2 shows the relationship between the GCE and other GGF working groups.

```
Portals/Apps/Users                     Grid Services

                                       |<--> GIS-WG
   Models-WG   <-->|                   |<--> Accts-WG
   Apps-WG     <-->|  <--> GCE-WG <--> |<--> Performance-WG
   Users-WG    <-->|                   |<--> Data-WG
   GCE-WG      <-->|                   |<--> Scheduling-WG
                                       |<--> Security-WG
```

**Figure 2:** The relationship between the GCE and other GGF groups.

In particular this group has identified workflow management as a promising means of coordinating tasks and activities within a GRID enabled environment propose to describe tasks, users, machines, workflows using meta-modeling techniques.

**The CMS ECAL CRISTAL Project**

CRISTAL is a scientific workflow system that is being used to control the production and assembly process of the CMS Electromagnetic Calorimeter (ECAL) detector at CERN Geneva [3]. The system has been used to coordinate and manage execution of the tasks and activities and the collections and compositions that occur in detector research and development. To enable flexibility and adaptability the CRISTAL system must be capable of describing and storing the tasks and activities of the domain to be automated, executing and co-ordinating the tasks and storing the outcomes. CRISTAL has taken an object-oriented approach describing all parts, manufacturing and production tasks and detector data using meta-modeling techniques [4]. One of the benefits of meta-modeling is that with careful analysis and use of descriptive classes a core software architecture can be developed to support almost any type of workflow system.

The first production version of CRISTAL has been in operation for almost two years and has gathered over 25 Gbytes of construction information for CMS ECAL. This version of CRISTAL employed meta-modeling techniques in what has come to be called a description-driven system [4]. Its technology was based on CORBA (Orbix) interfaces to C++ objects that track individual (versions) of detector elements and production tasks and record data in an Objectivity/DB repository. The current version CRISTAL-2, nearing delivery, is based on components that are implemented using Java and are fully described in a description-driven Berkeley/DB repository and consequently it benefits from a reusable and configurable design.

The next phase of development is an extension of the CRISTAL-2 work to provide a general purpose HEP enterprise environment for GRID users, called CRISTAL-G. Instead of using CORBA services CRISTAL-G will use GRID services such as MDS, GRAM and GASS [5]. It may be possible to leverage from current projects in the GCE research group such as CORBA CoG [6] which is intending to provide a CORBA toolkit to access GRID services. The first objective will be to provide GRID users with a distributed workflow management capability within which to describe multi-step tasks and activities to be performed across a GRID. The outcomes associated with the tasks and activities will be stored and managed by CRISTAL-G. The main functionality described here already exists so the main task will be porting the current CRISTAL system from CORBA to a GRID based environment. The system will use GRID protocols to discover, schedule and monitor distributed collaborative computational activities. However this requires collaboration of all the GRID entities involved. Therefore the second more difficult objective is to research ways of improving collaboration and discovery.

There are a number of possible ways forward. In the business and enterprise community one proposal is that each entity has a role then the model can be one of loosely coupled roles exchanging documents based on contract of collaboration. Role based models are favoured in business organisations but for a loosely coupled collaboration of institutes that occur in scientific communities the roles and models of interaction are less well defined. What is required is a mechanism so that these interactions and activities can be described and defined as a group is formed. Have to move to higher levels of abstraction to manage complexity (meta modelling) and to model this interaction. This is now well known and as a consequence there has been an explosion of meta-models within the object oriented software community. The CRISTAL group already has work in progress exploring the use of XML to exchange schemas and meta-models [7]. The descriptions of data and interactions will all be different so one way forward would be the semantically grounding of definitions as suggested by the DAML (+OIL) initiative of the DARPA community [8].

**CRISTAL-G and Information Grids**

As a consequence of mapping the description-driven approach of CRISTAL onto a data and computational GRID architecture it will be possible to have data which is logically separated and separately queryable from its interpretation. This approach of multiply layered systems has recently been pervasive in metacomputing, distributed systems, cooperative working, federated databases and knowledge management. It facilitates the provision of a so-called 'Information Grid' as has been advocated as the basis of a 3-layer architecture for grid in the UK's e-science initiative. Here the architecture comprises a Computation / Data Grid with networking as the base, an Information Grid for integrating heterogeneous information into an homogeneous presentation above and on top a Knowledge Grid for the extraction of knowledge from information and assists for user control of processes [9].

Currently there has been much activity in the US Grids community in the Computation/Data but little at the higher levels of data abstraction (with the notable exception of the semantic web [10]). Open research challenges at the Information Grid layer of implementation include, amongst others, the provision of appropriate software technologies to:

- Provide homogeneous access to heterogeneous resources on the Information Grid
- Handle the distribution and delivery of information and its content
- Provide 'active' content management (distribution, security etc)
- Facilitate the location of information and its description
- Implement middleware for information access and integration

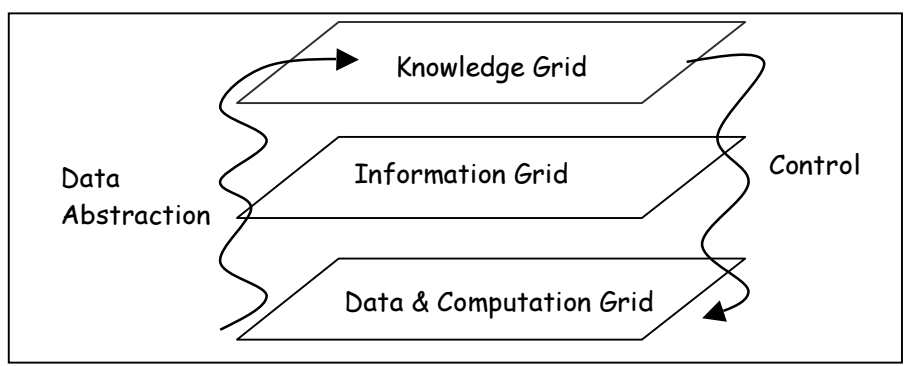

**Figure 3:** Three-layer Grid Architecture

Practitioners and researchers in the fields of distributed object-based systems, meta-data management, active middleware and reflective component-based architectures need to resolve these issues. What is required is reflective middleware (such as [10]) and middleware services which enable users of Grids, such as particle physics analysts, to execute queries which can use the *nature and content* of the Information Grid, including meta-data, to assist in locating data and resolving domain-related analyses. Mapping a description-driven system onto a GRID infrastructure is one step in achieving this.

**Conclusions and Future Work**

The focus of a GRID computing on large scale resource sharing and virtual organizations is certainly important to the scientific community. However as far as distributed computing is concerned it is only useful if it is able to progress beyond a framework for sharing a collection of distributed resources, something which has already been demonstrated at the computation level. The ideal concept of GRID computing requires an advance in interoperability at the collaboration level of knowledge and meta-model level interchange. CRISTAL and CRISTAL-2 have made considerable progress in advancing understanding of self-descriptive and meta-models in distributed HEP applications. The CRISTAL-G project is exploiting these ideas in the creation of a general problem solving environment which will allow GRID users to define, co-ordinate and manage GRID tasks and activities.